\def\rfr#1{eq. (\ref{#1})}

\def\leti{Lense--Thirring}
\def\Rfr#1{Eq. (\ref{#1})}

\def\derp#1#2{\rp{\partial{#1}}{\partial{#2}}}

\def\bar{\begin{eqnarray}}
\def\ear{\end{eqnarray}}

\def\eqi{\begin{equation}}
\def\eqf{\end{equation}}
\def\eqia{\begin{eqnarray}}
\def\eqfa{\end{eqnarray}}
\def\rp#1#2{{#1\over#2}}
\def\ct#1{\cite{#1}}
\def\lb#1{\label{#1}}

\def\oc2{$\mathcal{O}(c^{-2})$}

\documentclass[11pt]{article}
\usepackage{amsmath,amsthm,amscd,amssymb}
\usepackage{latexsym}
\usepackage{graphicx,epsfig}

\begin{document}

\noindent{\bf \LARGE{On the reliability of the so far performed
tests for measuring the Lense-Thirring effect with the LAGEOS
satellites}}
\\
\\
\\
{Lorenzo Iorio}\\
{\it Dipartimento Interateneo di Fisica dell' Universit${\rm
\grave{a}}$ di Bari
\\Via Amendola 173, 70126\\Bari, Italy
\\e-mail: lorenzo.iorio@libero.it}

\begin{abstract}
In this paper we critically discuss the so far performed attempts
aimed at the detection of the general relativistic gravitomagnetic
Lense-Thirring effect in the gravitational field of the Earth with
the existing LAGEOS satellites. In the latest reported measurement
of the gravitomagnetic shift with the nodes of the LAGEOS
satellites and the 2nd generation GRACE-only EIGEN-GRACE02S Earth
gravity model over an observational time span of 11 years a
5-10$\%$ total accuracy is claimed at 1-3$\sigma$, respectively.
We will show that, instead, it might be 15-$45\%$ (1-3$\sigma$) if
the impact of the secular variations of the even zonal harmonics
is considered. Possible strategies in order both to make more
robust and reliable the tests with the node-only LAGEOS-LAGEOS II
combination used and to overcome the problems affecting it with
other alternative combinations are presented.
\end{abstract}
\newpage
\tableofcontents
\newpage
\section{Introduction}
Recent years have seen increasing efforts aimed to
directly\footnote{According to K. Nordtvedt \ct{nor03}, the
multidecadal analysis of the Moon'orbit by means of the Lunar
Laser Ranging (LLR) technique yields a comprehensive test of the
various parts of order $\mathcal{O}(c^{-2})$ of the post-Newtonian
equation of motion. The existence of gravitomagnetism as predicted
by the Einstein's General Theory of Relativity would, then, be
indirectly inferred from the high accuracy of the lunar orbital
reconstruction. } detecting various phenomena connected to the
general relativistic gravitomagnetic field \ct{mash01, ciu95,
tart, scha} of the rotating Earth.

The extraordinarily sophisticated and expensive Gravity Probe B
(GP-B) mission \ct{eve, evetal01} has been launched in April 2004;
it is aimed at the detection of the gravitomagnetic precession of
the spins \ct{sch60} of four superconducting gyroscopes carried
onboard at a claimed accuracy of 1$\%$ or better.

The Lense-Thirring effect on the orbital motion of a test particle
\ct{lenti} could be measured by analyzing the orbital data of
certain Earth artificial satellites with the Satellite Laser
Ranging (SLR) technique \ct{NC02}.  Up to now, the only performed
tests are due to Ciufolini and coworkers \ct{science98, ciucaz04,
nature04, cazzarola}. In such papers a confirmation of the
existence of the Lense-Thirring effect as predicted by the
Einstein's General Theory of Relativity is claimed with a total
20-25$\%$ (node-node-perigee LAGEOS-LAGEOS II linear combination)
and $5-10\%$ (node-node LAGEOS-LAGEOS II linear combination)
accuracy, respectively, according to the adopted observable.

In this paper we will analyze the latest results presented in
\ct{ciucaz04, nature04, cazzarola} from a critical point of view
in order to show that the total error budgets might have been
underestimated.
\section[The Lense-Thirring effect on the orbit of a test particle]{The Lense-Thirring effect on the orbit of a test particle and the strategy to measure it}
\begin{figure}
\begin{center}
\includegraphics{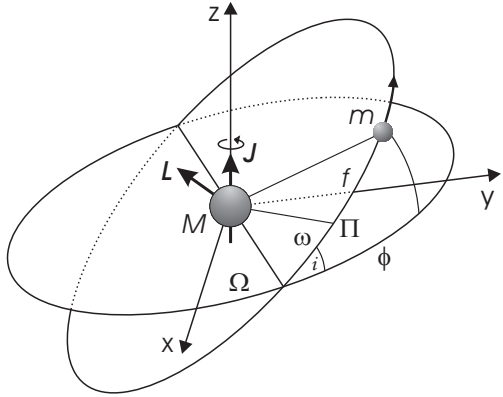}
\end{center}
\caption{\label{orbita}Orbital geometry for a motion around a
central mass. Here $L$ denotes the orbital angular momentum of the
particle of mass $m$, $J$ is the proper angular momentum of the
central mass $M$, $\Pi$ denotes the pericentre position, $f$ is
the true anomaly of $m$, which is counted from $\Pi$, $\Omega$,
$\omega$ and $i$ are the longitude of the ascending node, the
argument of pericentre and the inclination of the orbit with
respect to the inertial frame $\{x,y,z\}$ and the azimuthal angle
$\phi$ is the right ascension counted from the $x$ axis.}
\end{figure}

The gravitomagnetic field of a spinning mass of proper angular
momentum $J$ induces tiny secular precessions on the longitude of
the ascending node $\Omega$ and the argument of
pericentre\footnote{In their original paper Lense and Thirring use
the longitude of pericentre $\varpi=\Omega+\omega$. } $\omega$ of
a test particle \ct{lenti, ash, ciu95, ior01}
\eqi\dot\Omega_{\rm LT}=\rp{2GJ}{c^2 a^3 (1-e^2)^{3/2}},\
\dot\omega_{\rm LT}=-\rp{6GJ\cos i}{c^2 a^3
(1-e^2)^{3/2}},\lb{letprec}\eqf
where $G$ is the Newtonian constant of gravitation, $c$ is the
speed of light in vacuum, $a,e$ and $i$ are the semimajor axis,
the eccentricity and the inclination, respectively, of the test
particle's orbit. See Figure \ref{orbita} for the orbital geometry
of a Keplerian ellipse.
In the terrestrial space environment the gravitomagnetic
precessions are very small: for the spherically symmetric geodetic
SLR LAGEOS satellites, whose orbital parameters are listed in
Table \ref{para}, they amount to a few tens of milliarcseconds per
year (mas yr$^{-1}$ in the following)
{\small\begin{table}\caption{Orbital parameters of the existing
LAGEOS and LAGEOS II and of the proposed LARES and their
Lense-Thirring node precessions. }\label{para}

\begin{tabular}{lllll}
\noalign{\hrule height 1.5pt}

Satellite & $a$ (km) & $e$ & $i^{\circ}$ & $\dot\Omega_{\rm LT}$ (mas yr$^{-1}$)  \\

\hline

LAGEOS    &  12270    & 0.0045 &  110 & 31\\
LAGEOS II &  12163    & 0.0135 & 52.64 & 31.5\\
LARES    & 12270    & 0.04  & 70 & 31\\

\noalign{\hrule height 1.5pt}
\end{tabular}

\end{table}}

The extraction of the Lense--Thirring precessions from the orbit
data analysis is very difficult due to a host of competing
classical orbital perturbations of gravitational \ct{iortid01,
iorcelmec03, iorpav01, pavior02} and non-gravitational \ct{ves99,
luc01, luc02, luc03, luc04, lucetal04} origin which have various
temporal signatures and are often quite larger than the
relativistic signal of interest. The most insidious ones are the
perturbations which have the same temporal signature of the
Lense-Thirring precessions\footnote{Also the perturbations which
grow quadratically in time are, of course, very dangerous. Those
induced by the secular variations of the even zonal harmonics of
the Earth's geopotential fall in this category, as we will see in
detail in Section \ref{zonrat}. Harmonic time-dependent
perturbations with periods longer than the observational time span
may also be insidious because they would resemble superimposed
linear trends \ct{iortid01}.}, i.e. secular trends. Indeed,
whatever the length of the adopted observational time span $T_{\rm
obs}$ is, they cannot be fitted and removed from the time series
without removing the relativistic signal as well. Then, it is of
the utmost importance to assess as more accurately and reliably as
possible their aliasing impact on the measurement of the
Lense-Thirring effect.

It turns out that the perigees of the LAGEOS-like satellites are
severely  affected by the non-gravitational perturbations,
contrary to the nodes. Moreover, since the non--conservative
forces depend on the structure, the shape and the rotational
status of the satellite their correct modelling is not a trivial
task and, as we will see later, introduces large uncertainties in
the correct assessment of the error budget in some of the
performed gravitomagnetic tests.
\subsection{The gravitational error}
The even ($\ell=2,4,6...$) zonal ($m=0$) harmonic coefficients
$J_{\ell}$ of the multipolar expansion of the Earth's
gravitational potential, called geopotential, induce secular
precessions\footnote{Also the subtle non--gravitational
Yarkovsky-Rubincam force, which is due to the interaction of the
Earth's electromagnetic IR radiation with the physical structure
of the LAGEOS satellites, induces secular effects on their nodes
and perigees \ct{luc02}.} on the node, the perigee and the mean
anomaly of any near-Earth artificial satellite \ct{kau66} which,
of course, depend only on its orbital configuration and are
independent of its physical structure. Such aliasing effects are
many orders of magnitude larger than the Lense-Thirring
precessions; the precision with which the even zonal harmonics are
known in the currently available Earth gravity models \ct{jgm3,
egm96, grim5c1, eigen1s, eigen2, eigen3p, eigen-grace02s, ggm01,
cg01c} would yield errors amounting to a significant fraction of
the Lense-Thirring precessions or even larger.

Even more dangerous are the perturbations induced by the secular
variations of the low degree even zonal harmonics $\dot
J_{\ell},\ell=2,4,6$ \ct{eanes, cox}. Indeed, such perturbations
grow quadratically in time if the shifts in mas are considered and
linearly in time if the rates in mas yr$^{-1}$ are considered.
Their impact on the orbital elements of the LAGEOS satellites have
been worked out in \ct{iormor04}. It turns out that, by using the
results of \ct{eanes}, the errors induced by $\dot J_2$ would
amount to 8$\%$, 14$\%$ and $5.4\%$ for the nodes of LAGEOS and
LAGEOS II and the perigee of LAGEOS II, respectively, over an
observational time span $T_{\rm obs}$ of just one year at
1$-\sigma$ level. This clearly shows that it would be impossible
to analyze single orbital elements.

The time-dependent harmonic perturbations \ct{iortid01, iorpav01,
pavior02} are less dangerous because if their periods are shorter
than the adopted observational time span they can be fitted and
removed from the time series. The most insidious tidal
perturbation is that induced by the even zonal (055.565)
constituent which has a period of 18.6 years and whose nominal
impact on the orbital elements of the LAGEOS satellites amounts to
thousands of mas \ct{iortid01}. However, it turns out that it does
not affect the observables which have been adopted for the
performed Lense-Thirring tests because its main component is of
degree $\ell=2$ and order $m=0$ (see Section \ref{combosec}).

\subsubsection{The butterfly configuration}\lb{buttersec}
A possible way to cope with the even zonal gravitational
perturbations is represented by the use of a pair of satellites in
the so-called butterfly configuration. In it the orbital planes of
the two satellites are shifted 180$^{\circ}$ apart, all other
orbital parameters being equal. In this case the sum of the nodes
\ct{ciu86, ciu89} $\dot\Omega^{ (i)}+\dot\Omega^{(i+180^{\circ})}$
and the difference of the perigees \ct{iordif1, iordif2, iorluc}
$\dot\omega^{(i)}-\dot\omega^{ (i+180^{\circ})}$ would allow to
cancel exactly out the bias due to all the even zonal harmonics of
the geopotential\footnote{If totally passive satellites are
considered, the difference of the perigees would yield less
accurate results than the sum of the nodes \ct{iorluc}.}; the
gravitomagnetic precessions would, instead, add up. Indeed, the
Lense-Thirring node precessions are independent of $i$, while the
classical geopotential node precessions depend on a multiplicative
factor $\cos i$ common to all degrees $\ell$ and on sums of even
powers of $\sin i$ which are different for the various degrees
$\ell$ \ct{iorcelmec03}; the Lense-Thirring perigee precessions
depend on $\cos i$ while the classical geopotential perigee
precessions depend on even powers of $\cos i$ and $\sin i$
\ct{iorcelmec03}. The butterfly configuration cannot be realized
with the present-day existing Earth artificial satellites. In
\ct{ciu86} it was proposed to launch a LAGEOS-like satellite-the
former LAGEOS III which later became LARES\footnote{The
eccentricity of LARES would amount to $e_{\rm LARES }=0.04$ in
order to perform other tests of post-Newtonian gravity with the
perigee. } \ct{asi}-with the same orbital parameters of LAGEOS
apart from the inclination which should be supplementary. That
idea is still alive, although it has not yet been approved by any
national space agency or institution. Recently, it has been
proposed to adopt the LARES orbital configuration for the OPTIS
relativity mission \ct{OPTIS, optis, optis2} currently under
examination by the German Space Agency (DLR). It should be noted
that with a LAGEOS-LARES/OPTIS butterfly configuration the
difference of the perigees would not be a good observable because
of the small eccentricity of the LAGEOS orbit. The exact
cancellation of the classical precessions due to all the even
zonal harmonics of geopotential could be achieved only if the
LARES/OPTIS orbital parameters were exactly equal to those of
LAGEOS. This would pose severe restrictions in term of the quality
and, consequently, the cost of the rocket launcher to be used.
Indeed, the realistically obtainable precision with the originally
proposed sum of the nodes would be affected by the unavoidable
departures of the LARES/OPTIS orbital parameters from their
nominal values, at least to a certain extent. Also the fact that
the eccentricities of LAGEOS and LARES/OPTIS would differ by one
order of magnitude should be accounted for. Indeed, in these cases
a residual aliasing effect due to all the even zonal harmonics
$J_2, J_4, J_6...$ would still occur. These topics have been
investigated in \ct{iorlar}. A different observable involving also
the nodes of LAGEOS and LAGEOS II has been proposed in
\ct{iorlucciuf02} for the LARES mission. It would dramatically
reduce the dependence of the systematic error due to the even
zonal harmonics of the geopotential on the unavoidable orbital
injection errors allowing also for a reduction of the costs of the
mission. Moreover, as it will become clear later, also the impact
of the $\dot J_{\ell}$ would be greatly reduced. On the contrary,
the simple sum of the nodes would be affected by $\dot J_2, \dot
J_4, \dot J_6...$ which might not be negligible over a time span
many years long.
\subsubsection{The linear combination approach}\lb{combosec}
The problem of reducing the impact of the mismodeling in the even
zonal harmonics of the geopotential with the currently existing
satellites can be coped in the following way.

Let us suppose we have at our disposal N (N$>1$) time series of
the residuals of those Keplerian orbital elements which are
affected by the geopotential with secular precessions, i.e. the
node and the perigee: let them be $\psi^{\rm A},$ A=LAGEOS, LAGEOS
II, etc. Let us write explicitly down the expressions of the
observed residuals of the rates of those elements
$\delta\dot\psi^{\rm A}_{\rm obs}$ in terms of the Lense-Thirring
effect $\dot\psi_{\rm LT}^{\rm A}$, of N-1 mismodelled classical
secular precessions $\dot\psi_{.\ell}^{\rm A}\delta J_{\ell}$
induced by those even zonal harmonics whose impact on the
measurement of the gravitomagnetic effect is to be reduced and of
the remaining mismodelled phenomena $\Delta$ which affect the
chosen orbital element \eqi\delta\dot\psi_{\rm obs}^{\rm
A}=\dot\psi_{\rm LT}^{\rm A}\mu+\underset{{\rm N-1\ terms
}}{\sum}\dot\psi_{.\ell}^{\rm A }\delta J_{\ell}+\Delta^{\rm A},\
\underset{{\rm N}}{\underbrace{{\rm A=LAGEOS,\ LAGEOS\ II,...}}}
\lb{equaz}\eqf The parameter\footnote{It can be expressed in terms
of the PPN $\gamma$ parameter \ct{wil93} as $\mu =(1+\gamma)/2$.}
$\mu $ is equal to 1 in the General Theory of Relativity and 0 in
Newtonian mechanics. The coefficients $\dot\psi_{.\ell}^{\rm A}$
are defined as \eqi\dot\psi_{.\ell}=\derp{{\dot\psi}_{\rm
class}}{J_{\ell}}\eqf and have been explicitly worked out for the
node and the perigee up to degree $\ell=20$ in \ct{NC02,
iorcelmec03}; they depend on some physical parameters of the
central mass ($GM$ and the mean equatorial radius $R$) and on the
satellite's semimajor axis $a$, the eccentricity $e$ and the
inclination $i$. We can think about \rfr{equaz} as an algebraic
nonhomogeneuous linear system of N equations in N unknowns which
are $\mu $ and the N-1 $\delta J_{\ell}$: solving it with respect
to $\mu $ allows to obtain a linear combination of orbital
residuals which is independent of the chosen N-1 even zonal
harmonics. In general, the orbital elements employed are the nodes
and the perigees and the even zonal harmonics cancelled are the
first N-1 low-degree ones.

This approach is, in principle, very efficient in reducing the
impact of the systematic error of gravitational origin because all
the classical precessions induced by the static and time-dependent
parts of the chosen N-1 $J_{\ell}$ do not affect the combination
for the Lense-Thirring effect. Moreover, it is flexible because it
can be applied to all satellites independently of their orbital
configuration, contrary to the butterfly configuration in which
the cancellation of the even zonal harmonics can be achieved only
for supplementary orbital planes and identical orbital parameters.
This approach has been adopted, among other things, in
\ct{iorlucciuf02, optis} in order to make the requirements on the
LARES/OPTIS orbital configuration less stringent. Apart from the
first orbital element which enters the combination with 1, the
other elements are weighted by multiplicative coefficients
$c_i(a,e,i)\neq 1$ which are built up with $\dot\psi_{.\ell}$ and,
then, depend on the orbital elements of the considered satellites.
Their magnitude is very important with respect to the
non-gravitational perturbations, which in general are not
cancelled out by the outlined method, and to the other
time-dependent perturbations of gravitational origin which are
neither even nor zonal so that they affect the obtained
combination as well. Values smaller than 1 for the $c_i$
coefficients are, in general, preferable because they reduce the
impact of such uncancelled perturbations. It is important to note
that the order with which the orbital elements enter the
combination is important: indeed, while the systematic error due
to the even zonal harmonics of the geopotential remains unchanged
if the orbital elements of a combination are exchanged, the
coefficients $c_i$ do change and, consequently, also the
non-gravitational error. The best results are obtained by choosing
the highest altitude satellite as first one and by inserting the
other satellites in order of decreasing altitudes.

This method was explicitly adopted for the first time in
\ct{ciu96} with the nodes of the LAGEOS satellites and the perigee
of LAGEOS II. The obtained combination is
\eqi\delta\dot\Omega^{\rm LAGEOS }_{\rm obs
}+c_1\delta\dot\Omega^{\rm LAGEOS\ II}_{\rm obs
}+c_2\delta\dot\omega^{\rm LAGEOS\ II}_{\rm obs }\sim \mu
60.2,\lb{ciufform}\eqf where $c_1=0.304$, $c_2=-0.350$ and 60.2 is
the slope, in mas yr$^{-1}$, of the expected gravitomagnetic
linear trend. \Rfr{ciufform} is insensitive to the first two even
zonal harmonics $J_2$ and $J_4$. It has been used in
\ct{science98} when the level of accuracy of the JGM3 \ct{jgm3}
and EGM96 \ct{egm96} Earth gravity models, available at that time,
imposed the cancellation of the first two even zonal harmonics, at
least.

In view of the great improvements in the Earth gravity field
modelling with the CHAMP \ct{pav} and, especially, GRACE
\ct{grace} missions an extensive search for alternative
combinations has been subsequently performed \ct{iorimp, iormor04,
iorproc, jas, ves}. In \ct{iormor04, iorproc} the following
combination has been proposed\footnote{The possibility of using
only the nodes of the LAGEOS satellites in view of the
improvements in the Earth gravity models from GRACE has been
propsed for the first time in \ct{grace}, although without
quantitative details.} \eqi\delta\dot\Omega^{\rm LAGEOS }_{\rm obs
}+k_1\delta\dot\Omega^{\rm LAGEOS\ II}_{\rm obs}\sim \mu
48.2,\lb{iorform}\eqf where $k_1= 0.546$ and 48.2 is the slope, in
mas yr$^{-1}$, of the expected gravitomagnetic linear trend. It
has been adopted for the test performed in \ct{nature04} with the
2nd generation GRACE-only EIGEN-GRACE02S Earth gravity model
\ct{eigen-grace02s}. \Rfr{iorform} allows to cancel out the first
even zonal harmonic $J_2$.
\section{The performed Lense-Thirring tests with the LAGEOS satellites}
The only performed tests aimed at the detection of the
Lense-Thirring precessions of \rfr{letprec} in the gravitational
field of the Earth with the existing LAGEOS satellites have been
performed, up to now, by Ciufolini and coworkers. They have used
the node-node-perigee combination of \rfr{ciufform} \ct{science98,
ciucaz04} and the node-node combination of \rfr{iorform}
\ct{nature04}. In these works it has often been claimed
that ``[...] {\it the Lense-Thirring effect exists
and its experimental value,} [...]{\it , fully agrees with the
prediction of general relativity.}"

The main objections to the results presented in these works can be
summarized as follows
\begin{itemize}
  \item The authors have not performed really robust and reliable tests e.g. by varying
  the length of the adopted observational time span, running
  backward and forward the initial epoch of the analysis, varying
  the secular rates of the even zonal harmonics in order to check
  their impact over different time spans, using different Earth
  gravity models in order to obtain a scatter plot of the obtained
  results. Instead, it seems that the authors, for a given data set, have always used from time to time those Earth
  gravity models which yielded just the closest results to what it is
  a priori expected from the General Theory of Relativity.
  Moreover, the impact of  a priori `imprint' effects of the
  Lense-Thirring signature itself on the coefficients of the Earth gravity
  models used may have driven the outcome of the performed tests
  just towards the expected result
  \item The total error budget has probably been
  underestimated, especially the systematic error of gravitational
  origin. E.g., the impact of the secular variations of the even zonal harmonics of the geopotential,
  which may become a very limiting factor over time spans many years long as those used, has not been
  addressed in a realistic, reliable and satisfactorily manner. Moreover, in some cases it has been incorrectly
  calculated by summing in quadrature the various contributions to
  the gravitational error (static even zonal harmonics, tides, secular variations of the even zonal harmonics)
  which can, instead, hardly be considered as uncorrelated.
  Almost always 1$-\sigma$ results have been presented without any
  explicit indication of this fact.
  \item All the relevant works of other authors, in which many of these issues have been addressed, have always been
  consistently ignored
  \item The node-node combination of \rfr{iorform} has been repeatedly and
  explicitly presented as a proper own result of the authors with
  references to their works (ref. [6] of \ct{ciucaz04} and ref. 19 of \ct{nature04})
  which, instead, have nothing to do with \rfr{iorform}
 \end{itemize}

\subsection{The node-node-perigee tests}
The combination of \rfr{ciufform} has been analyzed by using the
EGM96 \ct{egm96} Earth gravity model over 4 years in
\ct{science98} and over 7.3 years in \ct{ciucaz04}. The claimed
total error budget amounts to 20-25$\%$ over 4 years and to 20$\%$
over 7.3 years.
\subsubsection{The gravitational error}
The impact of the remaining uncancelled even zonal harmonics of
the geopotential $J_6, J_8, J_{10},...$ on \rfr{ciufform} has been
estimated by Ciufolini and coworkers with the full covariance
matrix of EGM96 in a root-sum-square calculation. In
\ct{science98} and, six years later, in \ct{ciucaz04} it is
claimed to be $\lesssim 13\%$. Apart from the fact that this is a
$1-\sigma$ level estimate, in \ct{ries}, as later acknowledged in
a number of papers \ct{iorimp, iorcelmec03, iorlar, iormor04,
iorproc, ves}, the use of the full covariance matrix of EGM96 has
been questioned. Indeed, it has been noted that in the EGM96
solution the recovered even zonal harmonics are strongly
reciprocally correlated; it seems, e.g., that the 13$\%$ value for
the systematic error due to geopotential is due to a lucky
correlation between $J_6$ and $J_8$ which are not cancelled by
\rfr{ciufform}. The point is that, according to \ct{ries}, nothing
would assure that the covariance matrix of EGM96, which is based
on a multi--year average that spans the 1970, 1980 and early 1990
decades, would reflect the true correlations between the even
zonal harmonics during the particular time intervals of a few
years adopted in the analyses by Ciufolini and coworkers. Then, a
more conservative, although pessimistic, approach would be to
consider the sum of the absolute values of the errors due to the
single even zonal as representative of the systematic error
induced by our uncertainty in the terrestrial gravitational field
according to EGM96 \ct{iorproc, iormor04}. In this case we would
get a conservative upper bound of 83$\%$ (1-$\sigma$). If a
root-sum-square calculation is performed by neglecting the
correlations between the even zonals a 45$\%$ 1-$\sigma$ error is
obtained \ct{iorcelmec03, iorproc, iormor04, ves}.

\subsubsection{The non-gravitational error}
Another important class of systematic errors is given by the
non--gravitational perturbations which affect especially the
perigee of LAGEOS II. The main problem is that it turned out that
their interaction with the structure of LAGEOS II changes in time
due to unpredictable modifications in the physical properties of
the LAGEOS II surface (orbital perturbations of radiative origin,
e.g. the solar radiation pressure and the Earth albedo) and in the
evolution of the spin dynamics of LAGEOS II (orbital perturbations
of thermal origin induced by the interaction of the
electromagnetic radiation of solar and terrestrial origin with the
physical structure of the satellites, in particular with their
corner--cube retroreflectors). Moreover, such tiny but insidious
effects were not entirely modelled in the GEODYN II software at
the time of the analysis of \cite{science98, ciucaz04}, so that it
is not easy to correctly and reliably assess their impact on the
total error budget of the measurement performed during that
particular time span. According to the evaluations in
\cite{luc02}, the systematic error due to the non--gravitational
perturbations over a time span of 7 years amounts to almost
28$\%$. However, according to \cite{ries}, their impact on the
measurement of the Lense--Thirring effect with the nodes of LAGEOS
and LAGEOS II and the perigee of LAGEOS II is, in general, quite
difficult to be reliably assessed.

So, by adding quadratically the gravitational and
non--gravitational errors of \cite{luc02} we obtain for the
systematic uncertainty $\delta\mu ^{\rm systematic}\sim 54\%$ if
we assume a 45$\%$ error due to geopotential. The sum of the
absolute values of the errors due to gepotential added
quadratically with the non--gravitational perturbations would
yield a total systematic error of $\delta\mu ^{\rm
systematic}\sim$ 88$\%$. It must be noted that the latter estimate
is rather similar to those released in \cite{ries}. Note also that
they are 1-$\sigma$ evaluations. Moreover, it should be considered
that the perigee of LAGEOS II is also sensitive to the eclipses
effect on certain non--gravitational perturbations. Such features
are, generally, not accounted for in all such estimates. An
attempt can be found in \cite{ves99} in which the impact of the
eclipses on the effect of the direct solar radiation pressure on
the LAGEOS--LAGEOS II Lense--Thirring measurement has been
evaluated: it should amount to almost 10$\%$ over an observational
time span of 4 years.
\subsection{The node-node tests}
The situation might turn out to be even worse for the results
presented for the node-node combination of \rfr{iorform}. Such
observable only cancels out the gravitational bias of the first
even zonal harmonic $J_2$, but has the great advantage of
discarding the perigee of LAGEOS II and its insidious
non-gravitational perturbations.

It has explicitly been proposed for the first time in \ct{iorproc,
iormor04}, although the possibility of using a LAGEOS-LAGEOS II
node-only observable was presented for the first time in
\ct{grace} without quantitative details. In \ct{ciucaz04} it seems
that the author refers to it as a proper own result with the ref.
[6], i.e. \ct{science98} of the present paper. In \ct{nature04}
the authors, instead, explicitly attribute it to themselves with
ref. 19, i.e. \ct{ciu86} of the present paper. In \ct{cazzarola}
an unfair use also of \ct{ciu89, ciu96} is present.
\subsubsection{The
gravitational error}\lb{zonrat} In \ct{ciucaz04} the node-node
combination of \rfr{iorform} has been analyzed with the EIGEN2
\ct{eigen2} and GGM01 \ct{ggm01} Earth gravity models over a time
span of almost 10 years.

In \ct{iormor04} the impact
of the static part of the geopotential, according to the
CHAMP-only EIGEN2 Earth gravity model, is evaluated as $18\%$
(1-$\sigma$ root-sum-square covariance calculation), $22\%$
(1-$\sigma$ root-sum-square calculation) and $37\%$ (1-$\sigma$
upper bound). Ciufolini only reports 18$\%$ obtained with the full
covariance matrix of EIGEN2 for which the same remarks as for
EGM96 holds. Moreover, he seems to ignore that EIGEN2 is only
based on six months of data and that the released sigmas of the
even zonal harmonics of low degree, which are the most relevant in
this kind of analyses with the LAGEOS satellites, are rather
optimistic, as pointed out in \ct{eigen2} and acknowledged in
\ct{iormor04}. In regard to the GGM01 model, the covariance matrix
was not publicly released. Cufolini correctly presents a 19$\%$
which is the 1$-\sigma$ upper bound obtained in \ct{iormor04}.
However, GGM01 is only based on 111 days of data. The author's
claim ``We conclude, using the Earth gravity model EIGEN-2S, that
the Lense-Thirring effect exists and its experimental value,
$\mu=0.98\pm 0.18$, fully agrees with the prediction of general
relativity" seems  too optimistic.

In \ct{nature04} the authors use the 2nd generation GRACE-only
EIGEN-GRACE02S Earth gravity model \ct{eigen-grace02s}. Also in
this case the full covariance matrix was not available and the
authors correctly report a systematic error due to the even zonal
harmonics of 3$\%$ (root-sum-square calculation) and 4$\%$ (upper
bound) at 1-$\sigma$ level. The same results have been obtained in
\ct{iorgrace}.

Major problems may arise  when the authors show their $a\ priori$
error analysis for the time-dependent gravitational perturbations
(solar and lunar Earth tides, secular trends in the even zonal
harmonics of the Earth's field and other periodic variations in
the Earth's harmonics). Indeed, in \ct{nature04} they claim that,
over an observational time span of 11 years, their impact would
be
2$\%$. This evaluation is based on ref. 30  of
\ct{nature04} which refers to the WEBER-SAT/LARES INFN study; it
has nothing to do with the present node-only LAGEOS-LAGEOS II
combination. 
\subsubsection{The impact of the secular rates of the uncancelled
even zonal harmonics} Moreover, this estimate may turn out to be
optimistic because of the secular variations of the even zonal
harmonics\footnote{The problem of the secular variations of the
even zonal harmonics in post-Newtonian tests of gravity with
LAGEOS satellites has been quantitatively  addressed for the first
time in \ct{lucc03}. In regard to the \leti\ measurement with
\rfr{iorform}, it has been, perhaps, misunderstood in
\ct{iormor04}. } $\dot J_{\ell}$. Indeed, \rfr{iorform} allows to
cancel out $\dot J_2$, but, in principle, is sensitive to $\dot
J_4$, $\dot J_6$,..., as pointed out in \ct{iorgrace}. The
uncertainties in the $\dot J_{\ell}$ are still quite large: e.g.,
according to Table 1 of \ct{cox}, there is not yet even full
consensus on the sign of such rates. On the other hand, their
impact on the
\leti\ measurement grow linearly in time\footnote{For a possible
alternative combination which would cancel out the first three
even zonal harmonics along with their temporal variations see
\ct{jas, ves}.}. Indeed, the mismodelled shift, in mas, of
\rfr{iorform} due to the secular variations of the uncancelled
even  zonal harmonics can be written as \eqi
\sum_{\ell=2}\left(\dot\Omega_{.\ell}^{\rm LAGEOS}+k_1
\dot\Omega_{.\ell}^{\rm LAGEOS\ II }\right)\frac{\delta\dot
J_{\ell}}{2}T^2_{\rm obs},\lb{quadr}\eqf where the coefficients
$\dot\Omega_{.\ell}$ are $\partial \dot\Omega_{\rm class}/\partial
J_{\ell}$ and have explicitly been
 calculated up to degree $\ell=20$ in \ct{NC02, iorcelmec03}. It must
be divided by the gravitomagnetic shift, in mas, of \rfr{iorform}
over the same observational time span \eqi \left(\dot\Omega_{\rm
LT}^{\rm LAGEOS}+k_1 \dot\Omega_{\rm LT}^{\rm LAGEOS\ II} \right)
T_{\rm obs}=48.2\ {\rm mas\ yr^{-1}}\ T_{\rm obs}.\eqf
 By assuming $\delta\dot J_4=0.6\times 10^{-11}$
yr$^{-1}$ and $\delta\dot J_6=0.5\times 10^{-11}$ yr$^{-1}$
\cite{cox}, it turns out that the  percent error on the
combination \rfr{iorform} grows linearly with $T_{\rm obs}$ and
would amount to $1\%$ over one year at $1-\sigma$ level. This
means that, over 11 years, their impact might range from 11$\%$
(1-$\sigma$) to 33$\%$ (3-$\sigma$). These evaluations hold on the
assumption that $\dot J_4$ and $\dot J_6$ retain their constant
sign over the adopted observational time span. At present, there
is, in fact, evidence for an inversion only in $\dot J_2$ since
1998 \ct{cox}, but \rfr{iorform} is insensitive to it.
In
\ct{cazzarola} some explanations about the procedure followed in
the analysis of \ct{nature04} can be found. Basically, they are as
follows
\begin{itemize}
  \item The time series of the combined residuals of the nodes of LAGEOS and LAGEOS
  II, built up without the even zonal rates in the background reference model for the Earth gravity field,
  has been fitted with some harmonic
  signals, a linear trend and a parabolic signal from which a $\dot J_4^{\rm eff}\sim 1.5\times
  10^{-11}$ yr$^{-1}$ has been determined. Then, the so obtained $\dot J_4^{\rm
  eff}$ would have been inserted in the dynamical force models of the orbital processor
  in order to re-analyze the same data
  yielding a certain value of $\mu$ which we conventionally define as $\mu_{\dot J_{4}^{\rm eff}}$.
  \item The time series of the combined residuals of the nodes of LAGEOS and LAGEOS
  II, built up with the default value of the even zonal rate $\dot J_4^{\rm eff}=1.41\times 10^{-11}$ yr$^{-1}$
  in the EIGEN-GRACE02S model,
  has been fitted, again, with some harmonic
  signals, a linear trend and a parabolic signal yielding a certain value of
  $\mu$ which we conventionally define as $\mu_{\dot J_{4}^{\rm default}}$.
  \item The so obtained values $\mu_{\dot J_{4}^{\rm eff}}$ and $\mu_{\dot J_{4}^{\rm default}}$
  have been compared by finding a
  1$\%$ variation only. The author of \ct{cazzarola} writes `` [...] using the value $\dot J_4^{Effective}\cong 1.5*10^{-11}$
  for the LAGEOS satellites that we obtained from fitting the combined residuals [...] resulted in a change of the measured
   value of frame-dragging by about $1\%$ only with respect to the case of using $\dot J_4=1.41*10^{-11}$ [...]''
\end{itemize}
\subsubsection{Some possible criticisms to the outlined approach}
  The value  $\dot J_4^{\rm eff}\sim 1.5\times
  10^{-11}$ yr$^{-1}$ measured with the combination of
  \rfr{iorform} is affected not only by $\dot J_6$ and the other higher degree even zonal harmonics but, more
  importantly, by the Lense-Thirring signature itself. Indeed, the
  combination of \rfr{iorform} is designed in order to only disentangle
  $J_2$ and the Lense-Thirring effect. So, it is not admissible to
  use the so obtained $\dot J_4^{\rm eff}$, which is coupled by construction to the Lense-Thirring effect,
  in order to reliably and correctly measure the Lense-Thirring
  effect itself. A combination suitably designed in order to
  measure  $J_4$ independently of $J_2, J_6$ and the General Theory of Relativity has been
  proposed in \ct{Ioriozonals}
\subsubsection{Some suggestions to improve the reliability and the
robustness of the  discussed test}
\begin{itemize}
\item The values of $\dot J_{\ell}$ to be used in the dynamical force models of the orbital processors
should have been determined in a Lense-Thirring-free fashion in
order to avoid `imprinting' effects on the outcome of the
analysis. Indeed, they could easily drive it just towards  the
expected result. In regard to this point, suitably designed linear
combinations of the residuals of some orbital elements of the
existing SLR satellites could be used in order to disentangle,
e.g., $J_2$, $J_4$, $J_6$ and the gravitomagnetic force
  \item For a given observational time span and a given background reference Earth gravity model, two time
  series of the combined residuals, built up with and without the aforementioned $\dot
  J_{\ell}$ GTR-free values in the background reference models,
  should be analyzed
  The
  difference between the so obtained $\mu$ parameters can, then, be
  evaluated
  \item The aforementioned procedure should be repeated for
  various observational time spans in order to have a more robust
  assessment of the importance of the secular variations of the even zonal harmonics in the proposed
  measurement.
  Indeed, their systematic error on the Lense-Thirring effect grows linearly
  \item The aforementioned procedures should be repeated for
  various Earth gravity models in order to have a scatter plot
\end{itemize}
\subsubsection{On the correct evaluation of the systematic error of
gravitational origin} Another controversial point is that it is
unlikely that the various errors of gravitational origin can be
summed in a root-sum-square way because of the unavoidable
correlations between the various phenomena of gravitational
origin. Indeed, in the adopted EIGEN-GRACE02S model $\dot J_4$ and
$\dot J_6$ are not solved for: $\dot J_2$ and $\dot J_4$ have been
assumed fixed, while $\dot J_6$ is absent. This means that the
recovered $J_{\ell}$ are affected by such rates.  It would, then,
be more conservative to linearly add the errors induced by them.
In this case, the $(J_{\ell}^{(0)}-\dot J_{\ell})$ error would
range from 15$\%$ (4$\%$+11$\%$) at $1-\sigma$ level to 45$\%$
(12$\%$+33$\%$) at $3-\sigma$ level over a 11-years long
observational time span\footnote{The evaluations of \rfr{quadr}
have been used.
}. The so obtained global gravitational error can be added in
quadrature to the non-gravitational error. Even by assuming the
2$\%$ authors' estimate of the time-dependent part of the
gravitational error, the upper bound errors would be
$\sqrt{(4+2)^2+2^2}\%=6\%$ at 1-$\sigma$,
$\sqrt{(8+4)^2+4^2}\%=13\%$ at 2-$\sigma$ and
$\sqrt{(12+6)^2+6^2}\%=19\%$ at 3-$\sigma$. Instead, in
\ct{nature04} the authors add in quadrature the doubled error due
to the static part of the geopotential (the 2$\times 4\%$ value
obtained from the sum of the individual error terms), their
perhaps optimistic evaluation of the error due to the time
dependent part of the geopotential and the non-gravitational error
getting $\sqrt{8^2+4^2+4^2}\%=10\%$ at 2-$\sigma$. On the other
hand, when they calculate the 3-$\sigma$ upper bound it seems that
they triple the 3$\%$ error due to the static part of the
geopotential obtained with a root-sum-square calculation  and add
it in quadrature to the other (not tripled) errors getting
$\sqrt{9^2+2^2+2^2}\%\leq 10\%$ at 3-$\sigma$.
\subsubsection{The a priori `memory' effect of the Lense-Thirring
signature on the adopted Earth gravity model} The following point
seems also worthy of discussion. The recovered values of the even
zonal harmonics in the GRACE models like EIGEN-GRACE02S retain, in
principle, a sort of `imprint' or `memory' of the General Theory
of Relativity, which, in fact, has not been modelled in the
currently released GFZ-GRACE-based models (F. Flechtner, GFZ team,
private communication, 2004). This is certainly true for the Earth
gravity models of the pre-CHAMP and GRACE era obtained from
multidecadal laser ranging to the geodetic satellites of LAGEOS
type, as discussed in \ct{ciu96, ries, ves}. This feature is also
valid for, e.g. GRACE which recovers the low degree even zonal
harmonics from the tracking of both satellites by GPS and the
medium-high degree geopotential coefficients from the observed
intersatellite distance variations. Indeed, from \ct{cheng} it can
be noted that the variation equations for the
Satellite-to-Satellite Tracking (SST) range $\Delta\rho$ and range
rate $\Delta\dot\rho$ of GRACE can be written in terms of the
in-plane radial and, especially, along-track components $R,T, V_R,
V_T$ of the position and velocity vectors, respectively. In turns,
they can be expressed as functions of the perturbations in all the
six Keplerian orbital elements (see (10)-(11), (A4)-(A6),
(A14)-(A16) and (A28)-(A30) of \ct{cheng} ). Now, the
gravitomagnetic off-diagonal components of the spacetime metric
also induce short-periodic 1 cycle per revolution (1 cpr) effects
\ct{lenti, casotto, ash} on all the Keplerian orbital elements,
apart from the secular trends on the node and the pericentre. This
means that there is also  a Lense-Thirring signature in all the
other typical satellite and intersatellite observables like ranges
and range-rates. It is likely that it mainly affects the
low-degree even zonal harmonics. In this case, it might happen
that the results of \ct{ciucaz04, nature04}, obtained with the
$J_2-$free node-node LAGEOS-LAGEOS II combination, have been
driven close to the expected result, at least to a certain extent,
just by this a priori Lense-Thirring `imprint' on the Earth
gravity model used.

\subsubsection{The LARES mission}
Finally, it is hard to understand why the authors of \ct{nature04}
very often refer to the LAGEOS-LARES proposed experiment and to
the related simulations and error budgets. It is rather confusing
and misleading. As already explained, the LAGEOS-LAGEOS II
combination of \rfr{iorform} is, by construction, designed in
order to exactly cancel out the $J_2$ term with an approach which
can be applied to any orbital configuration given a pair of
satellites in different orbits or a pair of different Keplerian
orbital elements of the same satellite (see Section
\ref{combosec}). On the contrary, the observable originally
proposed for the LAGEOS-LARES mission is the simple sum of their
nodes. As pointed out in Section \ref{buttersec}, if the orbital
parameters of LARES, quoted in Table \ref{para}, were exactly
equal to their nominal values, all the even zonal harmonics would
be exactly cancelled out. Instead, the sum of the nodes would be
affected, to a certain extent, by the whole range of the even
zonal harmonics of the geopotential due to unavoidable departures
from the LARES nominal configuration because of the orbital
injection errors and mission design (the eccentricity of LARES
would be one order of magnitude larger than that of LAGEOS), i.e.
the coefficients of the classical nodal precessions would not be
exactly equal and opposite $\dot\Omega_{.\ell}^{\rm LAGEOS}\neq
-\dot\Omega_{.\ell}^{\rm LARES}$ for $\ell=2,4,6,8,...$. In
\ct{nature04} the combination of the nodes of LAGEOS and LAGEOS II
of \rfr{iorform} is presented as if it is only slightly different
with respect to the sum of the nodes of the originally proposed
LAGEOS-LARES configuration, apart from a 18$^{\circ}$ offset in
the inclination of LAGEOS II with respect to LARES. The
differences in the eccentricities and the semimajor axes, which do
play a role \ct{iorlar}, have been neglected.

\section{Conclusions}
In this paper we have performed a detailed critical analysis of
the reliability and robustness of the so far performed tests aimed
at the detection of the Lense-Thirring effect in the gravitational
field of the Earth with the existing or proposed LAGEOS
satellites.

We can summarize our conclusions as follows
\begin{itemize}
\item
In regard to the node-node-perigee LAGEOS-LAGEOS II combination,
still presented in recent works as \ct{ciucaz04}, the claimed
$20-25\%$ total accuracy obtained with the EGM96 Earth gravity
model seems to be too optimistic mainly because of an
underestimation of the systematic error due to the geopotential.
\item In regard to the node-node LAGEOS-LAGEOS II
combination of \rfr{iorform}, extensive and thorough tests are
required in order to check the impact of the secular variations of
the even zonal harmonics of the geopotential which may represent a
very limiting factor over time spans many years long. Such tests
should be conducted by varying the data sets, shifting backward
and forward the initial epoch of the analyses, varying the
magnitudes of the $\dot J_{\ell}$ in the force models for a given
observational time span and the length of the time span for given
values of $\dot J_{\ell}$. If it will turn out that $\dot J_4,\dot
J_6,... $ do affect \rfr{iorform}, only improvements in our
knowledge of the $\dot J_{\ell}$ will allow to use this observable
in a truly reliable and confident way; a better knowledge only of
the static part of the geopotential would not be sufficient.
Moreover, also the problem of the Lense-Thirring `memory' in the
adopted gravity model should be accounted for.
\item Alternative combinations involving the use of existing
SLR targets other than the LAGEOS satellites should be analyzed.
The most promising combination involves the nodes of LAGEOS,
LAGEOS II, Ajisai and Jason-1 \ct{jas, ves}. It cancels out the
first three even zonal harmonics $J_2, J_4, J_6$, along with their
temporal variations, at the price of introducing the relatively
huge  non-gravitational perturbations on Jason-1 which, however,
should have a time-dependent harmonic signature with short
periodicities. Moreover, in this case the possible impact of the
Lense-Thirring `imprint' would be greatly reduced. According to
the recently released combined CHAMP+GRACE+terrestrial
gravimetry/altimetry EIGEN-CG01C Earth gravity model \ct{cg01c},
the systematic error due to the remaining even zonal harmonics
would amount to 0.7 $\%$ (root-sum-square calculation) and 1.6$\%$
(upper bound) at 1$-\sigma$ \ct{jas}. The possibility of getting
long time series of the Jason's node should be seriously
investigated with real data tests
\item
The launch of a third LAGEOS-like satellite would allow for a
really robust and confident measurement of the Lense-Thirring
effect. Indeed, it would be possible to combine its node-and,
perhaps, also its perigee if it will be built up in such a way to
sufficiently reduce the impact of the non-gravitational
perturbations-with the nodes of the other currently existing
LAGEOS satellites in order to cancel out $J_2$, $J_4$ and,
perhaps, $J_6$ along with their temporal variations. The great
improvements in our knowledge of the terrestrial gravitational
field would allow to abandon the very tight and stringent
requirements of the originally proposed butterfly configuration.
However, since the LARES data should be combined with those of the
existing LAGEOS satellites, the limit of the total available
accuracy is mainly set by the impact of the non-gravitational
perturbations, i.e. it is of the order of $1\%$. The systematic
error of gravitational origin could be kept well below this value.
\end{itemize}

\section*{Acknowledgments}
I thank  H. Lichtenegger for Figure 1 and G. Melki for his useful
remarks.


\end{document}